\makeatletter

\declare@file@substitution{revtex4-1.cls}{revtex4-2.cls}
\makeatother
\documentclass[twocolumn]{aastex63}
\usepackage{epsfig}                     
\usepackage{graphicx}                    
\usepackage{courier}     
\usepackage{natbib}                 
\usepackage{amssymb}                    
\usepackage{color}                       
\usepackage{hyperref}
\usepackage[space]{grffile}
\usepackage{url}
\graphicspath{{./}{figures/}}

\begin{document}

\title{Cosmic Microwave Background Radiation within the Zwicky Tired Light Hypothesis}



\author{C.E. Navia}
\affiliation{Instituto de F\'{i}sica, Universidade Federal Fluminense, 24210-346, Niter\'{o}i, RJ, Brazil }
\correspondingauthor{C.E. Navia}
\email{carlos\_navia@id.uff.br}


\begin{abstract}
At high redshifts, the distribution of angular separations between galaxies in clusters is in tension with the standard cosmological model. The JWST data at even higher redshifts have exacerbated this tension. However, Zwick’s Tired Light (TL) model fits well with these high-redshift data, bringing the TL model back into the spotlight. In the TL model, redshift is attributed to the energy loss of photons during propagation. However, some argue that this hypothesis cannot explain the shape and primordial nature of the cosmic microwave background (CMB).
Assuming that the CMB’s source is the microwave radiation from galaxies, we show that the TL model can produce a CMB spectrum consistent with the COBE FIRAS data. The TL model predicts a finite optical path per Hubble radius, which defines an opaque region, a sphere with radius $R_{max}=240$ Mpc ($z=0.051$) centered on the Milky Way, 
within this region, the galaxy's microwave radiation superposition behaves like a black body.
 Additionally, the TL scenario (microwave radiation superposition) allows us to determine the angular size of the CMB temperature fluctuations, in agreement with the observed CMB power spectrum, and provides insight into the Sunyaev-Zeldovich effect.
\end{abstract}


\section{Introduction} 
\label{intro}

Hubble's (1929) observations of the redshift of light from galaxies, interpreted as the Doppler effect and its increase with distance from the galaxy, were anticipated by Lemaître's model of an expanding universe (Lemaître, 1927). Even Einstein conceded to this evidence, abandoning his model of a static universe and its cosmological constant. However, not everyone accepted the idea of an expanding universe. Some notable figures who disagreed included Fritz Zwicky, Max Born, Louis de Broglie, and Fred Hoyle
\cite{peck97,assi95}. For \cite{dira38}, the expansion of the universe is due to the gravitational constant, G, which changes over time in proportion (inversely) to the universe's age.

For \cite{zwic29}, the redshift observed in galaxies, interpreted as a Doppler effect indicating an expanding universe, would be merely an illusion. The Doppler effect does not provide any information about the nature of matter; it is only a kinematic issue. Zwicky proposed the ``tired light'' hypothesis, suggesting that the redshift originates from the loss of energy by photons during their journey.

\cite{finl54}, observing spectral lines of
B and O-type stars in the constellation of Orion, found
inconsistencies in the redshift values. Extrapolating to
the case of the cosmological redshift (as tired light) obtained
an intergalactic temperature of about $1.5^{\circ}K$, not
far from CMB’s temperature from COBE-FIRAS. In addition,
Eddington (1926) predicted a $\sim 3^{\circ}K$ for interstellar
temperature based on the radiation quantity from
stars and their distribution in space \citep{peck97,assi95}.

Later, in 1965, the accidental discovery of the CMB by Penzias and Wilson \citep{penz79}, along with the observation by \cite{dick65} that the CMB had already been predicted in 1948 by Gamow and colleagues \cite{gamo48,alph49} as a probable afterglow of the Big Bang, helped solidify the cosmological view of an expanding universe.

On the other hand, cosmological models were tested using the 1979 data on the angular size distribution of galaxies within clusters, compared to the clusters' redshifts. The data includes moderately high redshift clusters, i.e., $0.02<z<0.46$ \citep{hick79}. These results are in tension with Big Bang cosmology (the LCDM theory), while the TL model fits the data best, gaining further support \cite{lavi86}.
The galaxy angular size data has since improved. In addition to increasing the statistics, the dataset now includes high redshift clusters ($0.02 \leq z \leq 15$). The highest redshift values come from deep sky observations taken by the James Webb Space Telescope (JWST) \citep{lovy22}. Once again, the TL model best fits the data, considering only a static, Euclidean space without requiring ad hoc assumptions. Other alternative cosmological models, including LCDM, are discussed by \cite{lope22}.

Furthermore, according to Big Bang cosmology, also known as the concordance cosmological model, the primordial elements produced after the Big Bang were primarily hydrogen (the most abundant), a small fraction of helium, and a tiny fraction of lithium. This primordial Big Bang nucleosynthesis (BBN) is considered one of the key pieces of evidence for Big Bang cosmology, which began with the seminal 1948 paper by Alpher, Bethe, and Gamow \citep{alph48}.

Other elements, such as carbon, are thought to have been formed through nucleosynthesis inside stars and dispersed into space only after the stars exploded as supernovae. A timeline for the sequence of these processes indicates that they took place over timescales of at least 1.0 Gyr.

However, the JWST observed a small galaxy in deep space, known as GS-z12 \citep{deug24}, which was already structured and contained reasonable amounts of carbon, just 0.35 Gyr after the Big Bang. This marks the first detection of an element other than hydrogen in the deep universe. The fact that carbon appeared so early suggests that the first stars may have formed in a very different way than previously thought. This result is in tension with the concordance cosmological model, which is widely supported by the scientific community.

\cite{gupt23} refers to this as an "impossible early galaxy" and suggests that the universe could be twice as old as current estimates. His result comes from a hybrid model, which incorporates Zwicky's Tired Light (TL) model, allowing for a reinterpretation of redshift as a TL phenomenon rather than being solely due to expansion.

Furthermore, the observation that the difference in the redshift of galaxies rotating in opposite directions increases with their distance from Earth \citep{sham24} aligns with expectations from Zwicky's TL model.

However, some argue that Zwicky's Tired Light (TL) model within a static Euclidean universe cannot explain the shape and origin of the CMB. In this work, we show that the TL model is consistent with the existence of an opaque region in the sky, where the optical depth per Hubble radius is finite (($y_{max}=0.05$), corresponding to a redshift of $z=0.051$. Within this region, the galaxy's microwave radiation superposition behaves like a blackbody. The TL model requires only two free parameters, one of which we determine by fitting the Hubble diagram of supernovae with the TL model.

This scenario also allows us to calculate the angular size of the CMB temperature fluctuations, which are consistent with the sizes observed in the CMB's power spectrum. We further provide a preliminary description of the Sunyaev-Zeldovich effect. Additionally, the observed anomalies in the CMB can be explained within this framework.

\section{Background}
\label{background}

Zwicky's tired light theory posits that the redshift observed in the light from galaxies is due to the loss of energy by photons during their cosmological propagation. The rate of photon energy loss per unit length is proportional to the photon's energy.

\begin{equation}
\frac{\delta E}{\delta d} \propto E.
\end{equation}.
this means
\begin{equation}
E = E_0 \exp\left({-\frac{d}{R_0}}\right),
\end{equation}
where $R_0$ is a free parameter.

Taking into account that  $E=h\nu$ where $h$ is Planck’s constant and $\nu$ is the frequency, along with the redshift defined as

\begin{equation}
(z+1)=\frac{\nu_0}{\nu}, 
\end{equation}
yields
\begin{equation}
(1+z) =\exp(d/R_0).
\label{main1}
\end{equation}

The above expression can mimic Hubble's law by considering
 $d<<R_0$ and expanding the exponential factor, in series ($\exp(d/R_0)\sim 1+d/R_0$),  yielding
\begin{equation}
z=\frac{d}{R_0},\;\;\;\;cz=H\;d,
\label{main2}
\end{equation}
where $R_0=c/H$ is the so-called Hubble radius and $H$ the Hubble constant. The main equation of TL model (Eq.~\ref{main1}) can be rewrite as
\begin{equation}
(1+z) =\exp\left[(H/c)\;d\right].
\label{main3}
\end{equation}

$R_0=c/H$ (Hubble radius)  is a free parameter in the tired light (TL) model. Its value can be obtained, for example, by fitting the supernova Hubble diagram data to the TL model prediction.
 
\subsection{Zwicky's Gravitational Drag Hypothesis}
\label{drag}

Zwicky \citep{krag17} discussed several possible explanations for the spectral shifts following the postulation of the tired light mechanism, ranging from Compton scattering of photons by free electrons to photon gravitational drag. The first explanation is inadequate because Compton scattering only occurs with high-energy photons (such as X-rays). The second explanation considers the gravitational mass of photons ($h\nu/c^2$) and, therefore, should be able to transfer momentum and energy to matter. Consequently, the gravitational drag hypothesis suggests that redshift should depend on the distribution of matter in space, meaning that redshift should vary with direction.

 This hypothesis was tested by Broughton (2014), who studied the radial velocities of globular clusters under the assumption of gravitational drag \citep{krag17}. Broughton concluded that the drag hypothesis is satisfactory.

\cite{land05} observed evidence for a preferred axis in the cosmic radiation anisotropy, known as the ``axis of evil'' an anomaly in the data from NASA’s Wilkinson Microwave Anisotropy Probe (WMAP) and confirmed by Planck detector, which shows anisotropic variations in the CMB radiation aligned with local structures (see section \ref{anomalies}). In other words, the gravitational drag effect on CMB photons caused by large concentrations of matter, such as the Virgo Cluster, could explain this anomaly.

\section{Hubble diagram}
\label{hubble}

Within the TL scenario, the radial distance to an object is determined exclusively by its redshift (see section~\ref{background}) and given by
\begin{equation}
d= R_0\; ln(1+z),
\end{equation}
where $R_0=c/H$ is only a free parameter, called Hubble radius.

The luminosity distance is related to the so-called ``comoving transverse distance" $d_M$ by
\begin{equation}
d_L=(1+z) d_M.
\end{equation}
In the case of a flat universe,  the comoving transverse distance $d_M$ is exactly equal to the radial comoving distance $d$ \citep{gabr04}. So, the relation between the luminosity distance $d_L$ and the radial distance $d$ is given as 
\begin{equation}
d_L=(1+z) d.
\end{equation}

Yielding the relation
\begin{equation}
d_L=(1+z)R_0\; ln(1+z),
\end{equation}

Furthermore, by expressing the brightness of an object in terms of its absolute magnitude 
M and considering standard candle-type objects, such as Type Ia supernovae, which have nearly the same absolute magnitude 
M, along with the apparent magnitude 
m, the distance to the object can be computed using the inverse square law. The distance modulus $\mu_0$ is defined as

\begin{equation}
\mu_0=m-M=5\;log_{10}(\frac{d_L}{Mpc}) +25.
\label{modulus}
\end{equation}
where $d_L$ is the luminosity distance. 

The Hubble diagram (HD) is a graph that plots the distance modulus against the redshift.

\begin{figure}[]
\vspace*{-0.0cm}
\hspace*{-0.5cm}
\centering
\includegraphics[clip,width=0.5
\textwidth,height=0.6\textheight,angle=0.] {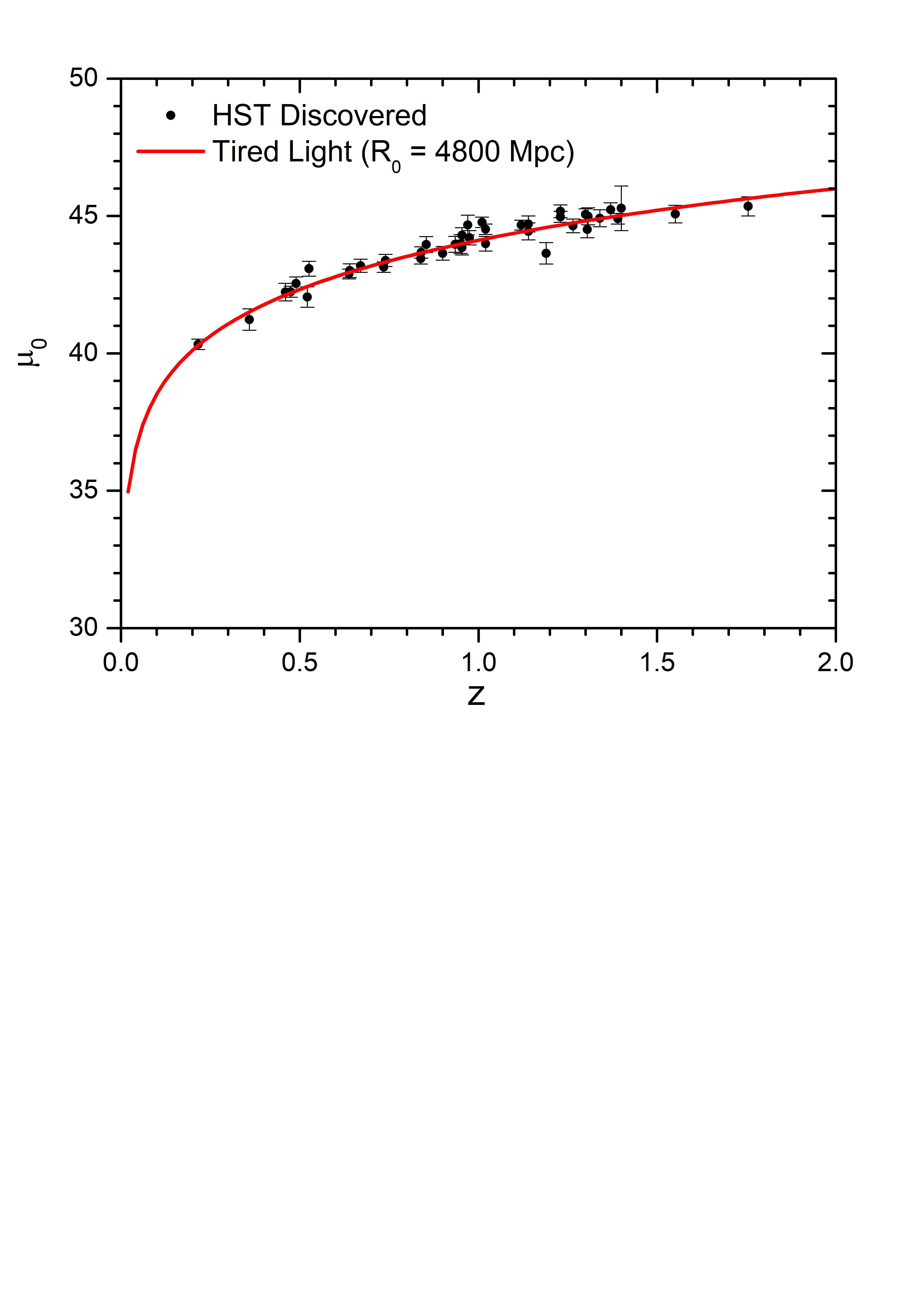}
\vspace*{-6.5cm}
\caption{Supernovae Hubble diagram, the black circles represent the Hubble Space Telescope data \citep{ries07}. The red curve is the TL model prediction to a Hubble radius $R_0=4800$ Mpc. 
}
\label{hubble1}
\end{figure} 

The Hubble Space Telescope (HST) supernova data \citep{ries07} were used to plot the Hubble diagram, as shown in Fig.~\ref{hubble1}. The black solid circles represent the HST supernova data, and the solid red curve represents the TL prediction, which fits the data with a single free parameter, $R_0 =4800$ Mpc. This mimics a Hubble constant 
$H=c/R_0=62.5$ km s$^{-1}$ Mpc$^{-1}$.

The supernova Hubble diagram was initially fitted by the 
LCDM model. However, this model requires at least two free parameters: the dark energy density and the cold dark matter density in the universe.

\begin{figure}[]
\vspace*{-0.0cm}
\hspace*{-1.6cm}
\centering
\includegraphics[clip,width=0.6
\textwidth,height=0.34\textheight,angle=0.] {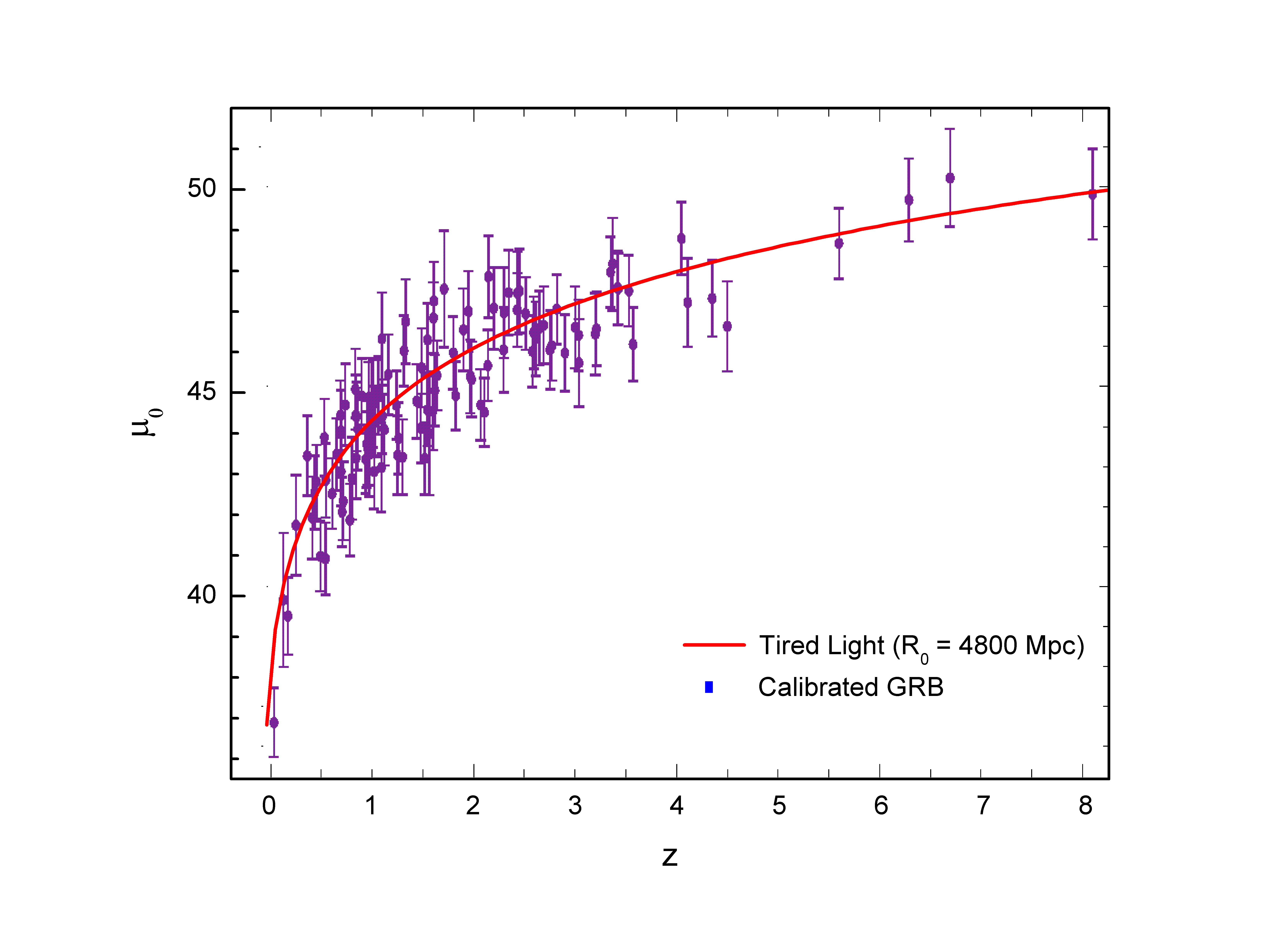}
\vspace*{-0.5cm}
\caption{GRB Hubble diagram, the blue points represent the GRB calibrated data \citep{demi12}. The red curve is the TL model prediction to to a Hubble radius $R_0=4800$ Mpc.
 }
\label{hubble2}
\end{figure} 

An extension of the Hubble diagram is possible up to redshifts around $z \sim 8$ through Gamma-Ray Bursts (GRBs). However, GRBs are not standard candles, as their luminosity peaks span a wide range. Nevertheless, it is possible to extract model-independent information from GRBs.

In some cases, the GRB Hubble diagram can be constructed from a sample of GRBs using only data from their X-ray afterglow light curve, specifically the break time of the X-ray beam. A sample of 66 calibrated Swift GRBs, using the Type Ia supernovae data \citep{demi12}, along with a data sample of 109 high-redshift GRBs calibrated from the Amati correlation \citep{demi12}, was used to construct the Hubble diagram shown in Fig.~\ref{hubble2}. Again, the red solid curve represents the TL model prediction with 
$R_0=4800$ Mpc.

\section{An alternative model for the origin of the CMB}
\label{radiance}

The most powerful evidence for the Big Bang comes from the CMB radiation. However, the CME, as a cosmological signal from the early universe, requires the assumption of two entities that are not yet fully understood: dark energy and cold dark matter. The latter has been the most sought-after entity, with negative results thus far, including experiments at the LHC particle accelerator at CERN. The LHC results have confirmed the standard model of particle physics \citep{azzi19}, with no evidence of exotic particles, such as supersymmetric particles, of which the ``neutralino'' is the leading candidate for dark matter.

The CMB radiation has been one of the main observables in cosmology since its accidental discovery by Penzias and Wilson \citep{penz79} in 1965. In the same year, Dicke, Peebles, and Wilkinson \citep{dick65} suggested that this radiation was predicted by Alpher and Gamow \citep{alph48} as fossil radiation, an afterglow of the Big Bang. From that moment on, the Big Bang theory became the predominant cosmological model, overshadowing its main competitor, the Steady State Theory \citep{bond48}, which had a respectable group of followers.

The fit of the standard cosmological model, LCDM, to the observed CMB power spectrum requires six free parameters, three of which are the densities of dark energy, cold dark matter, and the Hubble parameter. The Hubble parameter has also been measured directly, primarily using supernova data.
As of August 2024, the combined JWST and HST data provide a value for the Hubble constant that differs by approximately 5 sigma from the value obtained by fitting LCDM to the CMB data \citep{ries24}. This discrepancy is known as the "Hubble tension." The recent discovery by JWST of the gravitationally lensed Supernova Hope further confirms this tension.


Here, we describe a toy model for obtaining the CMB spectrum within Zwicky's TL model. We assume that the CMB is produced by the radiation of galaxies within a sphere of radius 
$R_{max}$, which is the only free parameter of the TL model. Starlight from galaxies loses energy during propagation, shifting to lower frequencies, in accordance with the TL scheme.

The Sloan Digital Sky Survey (SDSS) star spectra range from approximately 100 nm (deep ultraviolet) to around 10,000 nm (far infrared), with a microwave component at the tail. A star's spectrum includes several dark absorption lines, produced by atoms whose electrons absorb light at specific wavelengths. However, the smooth curve of an actual star's spectrum closely approximates the ideal thermal stellar spectrum \citep{sloan4}, which aligns reasonably well with a blackbody radiator. This is the case for the Sun, with a temperature of $T=5700^{\circ}K$. Additionally, 17 stars in the SDSS have spectra that are very close to blackbody radiation \citep{suzu18}. Approximately 99\% of the Sun's electromagnetic radiation falls within the ultraviolet, visible, and infrared regions, with less than 1\% in the microwave region.

In addition, galaxies such as the Milky Way contain at least 100 billion stars, with millions of galaxies contributing to the CMB. The superposition of these spectra, each with different temperatures, propagating over large distances (on the order of kpc and Mpc), results in photons being scattered multiple times and thermalized, ultimately forming a nearly isotropic spectrum. Based on the radiation from stars and their distribution in space, Eddington predicted in 1926 an interstellar temperature of 
$3^{\circ}K$.

Let’s begin by considering Planck's law for the spectral radiance of a black body

\begin{equation}
B_{\nu}(T,\nu)=\frac{2h\nu^3}{c^2}\times \frac{1}{e^{h\nu/kT}-1},
\label{planck}
\end{equation}
where $\nu$ is the frequency, h is the Planck constant, k is the Boltzmann constant, 
c is the speed of light, and T is the temperature.
The above relation describes the spectral radiance of the CMB with a temperature 
$T=T_0=2.73^{\circ}$K.

The spectral radiance as a function of redshift can be obtained by considering the evolution of the frequency.
If $\nu$ is the emission frequency from a source with redshift z, the observed frequency will be
\begin{equation}
\nu_{obs}= \frac{\nu}{ (1+z)},
\end{equation}

and the spectral radiance becomes

\begin{equation}
B_{\nu}(\nu,z)=\frac{2h\nu^3}{c^2}\times \frac{1}{e^{h\nu/kT_0(1+z)}-1}\times (1+z)^{-3}.
\label{blackbody_tl}
\end{equation} 

In the above expression, $T_0$ represents the temperature at 
$z=0$, and we assume that  $B_{\nu}(\nu,z)$
 is the spectral radiance emitted by a galaxy at redshift z.
 
 To account for N galaxies, each with redshift $z_i$, the contribution of these 
N galaxies will be

\begin{equation}
B_{\nu}(\nu)= \sum_{i=1}^{i=N} B_{\nu}(\nu,z_i).
\label{summatory}
\end{equation} 

The number of galaxies (N) contributing to the CMB's spectral radiance is calculated by assuming that, on large scales, the galaxy distribution is isotropic. If n denotes the galaxy density and A represents the effective area of a galaxy emitting radiation, the number of galaxies, dN, within a spherical shell of thickness dr, can be written as

\begin{equation}
dN= nA dr.
\end{equation} 

The contribution of N galaxies within a sphere of radius $R_{max}$ to the CMB spectrum is obtained by integrating the spectral radiance between the limits $r=0$
 and $r=R_{max}$. This requires a change of variable from z to the radial distance 
r. In the case of the TL model, we have

\begin{equation} 
(1+z)= \exp (r/R_0)=\exp (y),
\end{equation}
\begin{equation}
(1+z)^{-3}= \exp (-3y),
\end{equation}
\begin{equation}
dr=R_0\;dy.
\end{equation}
\begin{equation}
dN=nAR_0\;dy=NA\;c/H.
\end{equation}

Consequently, the summation in Eq.~\ref{summatory} becomes an integral

\begin{equation}
B_{\nu}(\nu)=\frac{1}{y_{max}} \times \frac{2h \nu^3}{c^2}.
 \int_{0}^{y_{max}} 
\frac{\exp(-3y)}{\exp(h/(kT_0)e^{-y})-1} dy,
\label{TL_radiance}
\end{equation} 

with $y_{max}^{-1}=nAR_0$, this constitutes the only free parameter of the TL model. 

\begin{figure}[]
\vspace*{-0.0cm}
\hspace*{-0.7cm}
\centering
\includegraphics[clip,width=0.45
\textwidth,height=0.6\textheight,angle=0.] {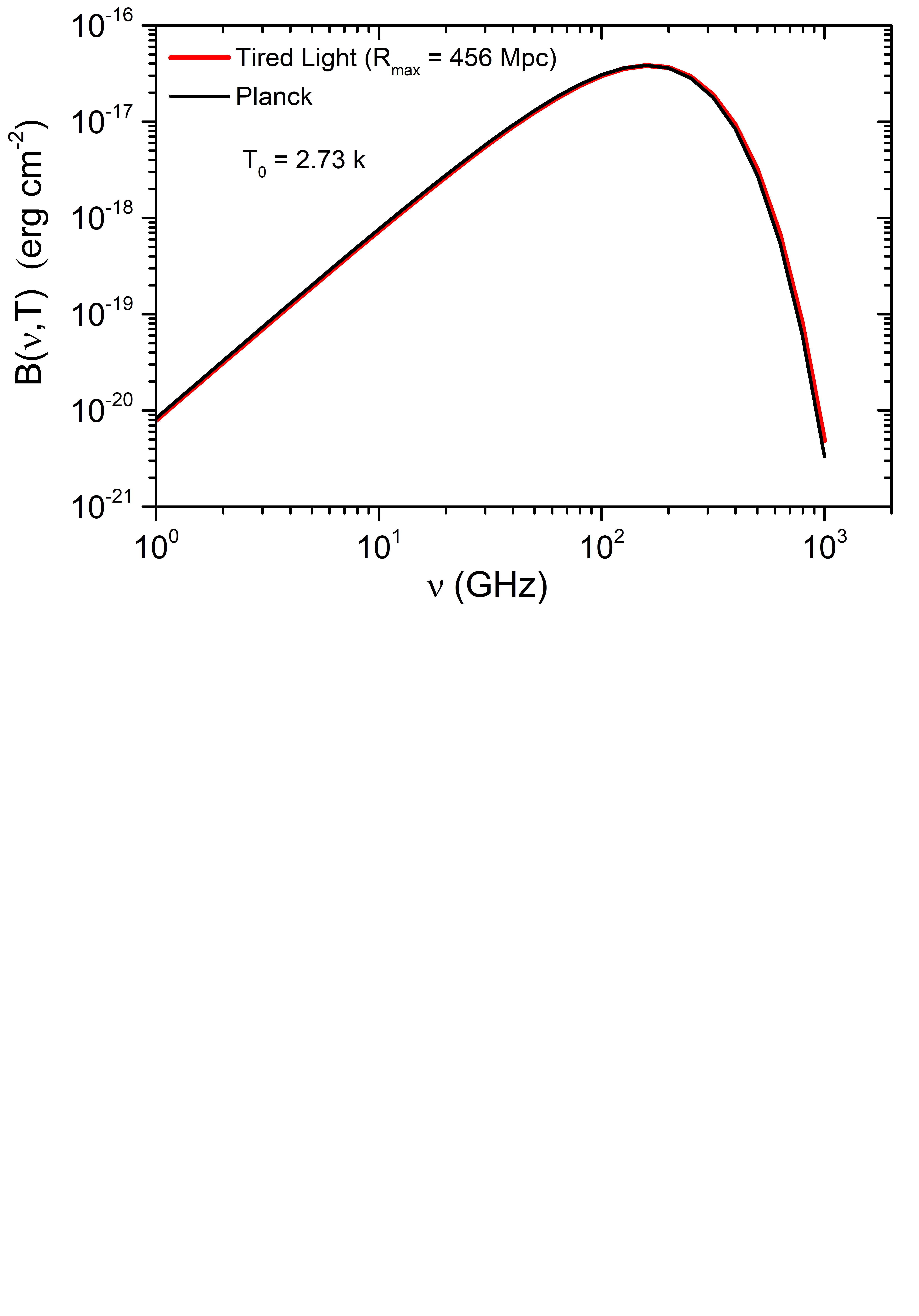}
\vspace*{-7.0cm}
\caption{Spectral radiance against frequency to a temperature of 2.725 k. The black line represents the Planck relation for a black body. The red line represents the TL model prediction. Only the microwave radiation emitted from galaxies within a radius $R_{max}=456$ Mpc contributes to the observed CMB radiation.
}
\label{radiance2}
\end{figure}

Fig.~\ref{radiance2} (red curve) shows the CMB’s spectral radiance to
 $T=2.73^{\circ}$K expected by the TL model, expressed by Eq.~\ref{TL_radiance}. It reproduces the blackbody spectral radiance from Eq.~\ref{planck} (black curve) only for the value $y_{max}=0.05$. 

\begin{figure}[]
\vspace*{-0.0cm}
\hspace*{-0.3cm}
\centering
\includegraphics[clip,width=0.45
\textwidth,height=0.6\textheight,angle=0.] {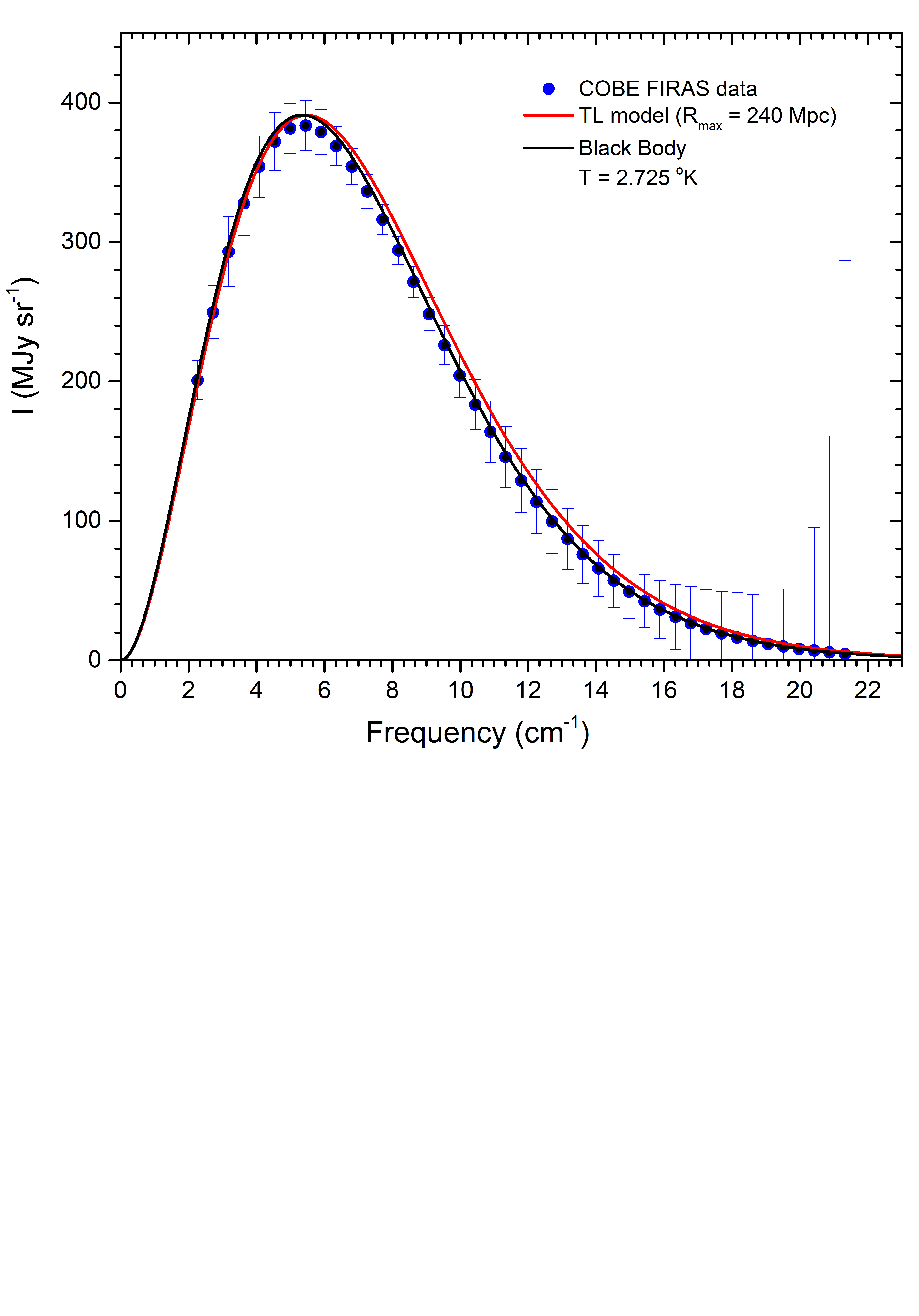}
\vspace*{-6.0cm}
\caption{Same as in Fig.~\ref{radiance2}, where the CMB's intensity replaces the spectral radiance, including a comparison with COBE FIRAS data.
}
\label{COBE_FIRAS}
\end{figure}

Already
Fig.~\ref{COBE_FIRAS} shows the CMB's intensity according to the  COBE-FIRAS data (black solid circles).  
The black curve represents a fit in the data assuming a black body intensity for $T_0=2.735\;^{\circ}K$, and the red curve expected by the TL model.

In the present case, the optical depth per Hubble radius is given by  
$y_{max}=(nAc/H)^{-1}=0.05$, resulting in an optical path $R_{max}=y_{max}\;c/H=240$ Mpc (z=0.051). This defines an opaque region, a sphere with radius $R_{max}$
centered on the Milky Way). Within this region, the galaxy's microwave radiation superposition behaves like a blackbody. The long journey of these photons (spanning distances from the order of kpc to hundreds of Mpc) causes them to lose up to 5\% of their energy and undergo multiple scatterings, contributing to their thermalization and isotropization. Therefore, the microwaves emitted from this opaque region resemble blackbody radiation.

The TL model also estimates the number of galaxies $N_{max}$ within the opaque region. Considering $y_{max}=(nAc/H)^{-1}=0.05$ and
$n=4/3 \;\pi R_{max}^3\;N_{max}$ we obtain:
$N_{max} \approx 10-30$ million galaxies, within $R_{max}=240$ Mpc (z=0.051).

\begin{figure}[]
\vspace*{-0.0cm}
\hspace*{-0.2cm}
\centering
\includegraphics[clip,width=0.45
\textwidth,height=0.35\textheight,angle=0.] {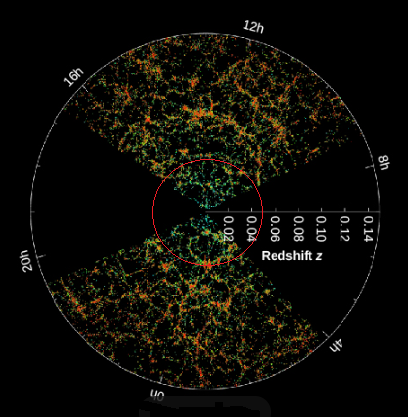}
\vspace*{-0.0cm}
\caption{SDSS 3-dimensional map of galaxy distribution. Earth is at the center, and each point represents a galaxy. The red one indicates galaxies with old stars. 
The galaxy sample has a median redshift of $z=0.1$.
According to the TL model,  the red circle with a radius z=0.051 indicates an opaque region. The radiation from these galaxies is like a microwave blackbody radiator.
Credit: Sloan Digital Sky Survey.
}
\label{sloan2}
\end{figure}

Fig.~\ref{sloan2} shows the Sloan Digital Sky Survey (SDSS) galaxy sky map, where each point represents a galaxy. According to the TL model, the opaque region emitting microwaves like a blackbody is contained within the red circle, which has a radius corresponding to $z=0.051$
 and is centered on the Milky Way. This is equivalent to an optical path per Hubble radius of $y_{max}=0.05$.

\section{Angular size of the CM fluctuations}
The CMB data's power spectrum shows that the CMB temperature oscillates with a frequency close to that of sound waves. Additionally, the power spectrum allows us to measure the angular size of the CMB temperature fluctuations.

According to many astrophysicists, the properties of the CMB cannot be described conventionally—i.e., without dark matter. However, in 2021, \cite{skor21} used a vector field and scalar field to mimic dark matter in a way similar to MOND's theory and were able to describe the CMB's power spectrum.

Nevertheless, the CMB's spectrum is predominantly analyzed within the LCDM model in order to derive cosmological parameters, such as the amounts of visible and dark matter and dark energy, that were present in the early universe.

In section~\ref{radiance}, we have shown that the CMB spectrum admits a new scenario within the TL model, suggesting an alternative origin for the CMB. In the TL framework, the CMB originates from galaxies emitting microwave radiation. However, the effective contribution comes only from galaxies with redshifts up to $z = 0.051$.

Due to the density fluctuations in the intergalactic medium during the CMB's propagation, the photon radiation density oscillates, producing regions of both low and high temperatures. Here, we present an alternative method for measuring the sizes of the CMB fluctuations using a toy model based on the TL scenario.

For simplicity, let us consider the oscillation shape of the photon radiation density as a solution to a damped harmonic oscillator.
\begin{equation}
\rho(t)=\rho_0 e^{-\gamma/2\;t} \cos(\omega t-\delta).
\end{equation}
with
\begin{equation}
\omega=\omega_0 \left(1-\frac{\gamma^2}{4\omega_0^2}\right)^{1/2},
\end{equation}
where $\omega_0$ is the undamped frequency, $\gamma$ the damped rate and $\delta$ is a constant which depend upon the initial conditions.

The strength of the oscillating radiation density, to 
$\gamma < \omega$ is
\begin{equation}
P(t) \sim P_0 e^{-\gamma t}\sin^2(\omega t-\delta),
\label{power2}
\end{equation}. 

If $\omega$ is the oscillation frequency for a  redshift z, the observed frequency will be
\begin{equation}
\omega_{obs}= \frac{\omega}{(1+z)},
\end{equation}
and the strength of the oscillating radiation density as a function of redshift becomes
\begin{equation}
P(t,z) \sim P_0 e^{-\gamma t}\sin^2(\omega /(z+1)\;t-\delta).
\label{power3}
\end{equation}
We assume that $P(t,z)$ is the strength of the oscillating radiation density due to a galaxy at redshift z.
The contribution of N galaxies, each with a redshift $z_i$, will be a summation of these N galaxies
\begin{equation}
P(t)= \sum_{i=1}^{i=N} P(t,z_i).
\label{summatory2}
\end{equation} 

Hereafter, the contribution of the number of galaxies (N) to the strength of the radiation density follows the same scheme presented in section~\ref{radiance}, with the consideration that the contributing galaxies are within a sphere of radius $R_{max}$. The number of galaxies, dN, within a spherical shell of thickness dr is $dN=nA\;dr$, where $n$ is the galaxy density, and A is the effective area of the galaxies emitting radiation. Additionally, according to the TL model  $(1+z)=\exp(d/R_0)$, and by defining $y=d/R_0$, 
the sum expressed in Eq.~\ref{summatory2} can be rewritten as an integral.

\begin{equation}
P(t) \sim (nA\;c/H) P_0 e^{-\gamma t} \int_{0}^{y{max}} \sin^2[\omega \exp(-y) t-\delta]dy,
\label{power4}
\end{equation}
With $y_{max}=(nA\;c/H)^{-1}$. Again, Eq..~\ref{power4} coincides with Eq.~\ref{power2} only for $y_{max}=0.05$. This implies that the superposition of damped microwave density oscillations within the opaque region ($z=0.051$) behaves like a single damped microwave density oscillation, i.e., the CME's damped oscillations.


\begin{figure}[]
\vspace*{-0.0cm}
\hspace*{-0.3cm}
\centering
\includegraphics[clip,width=0.45
\textwidth,height=0.45\textheight,angle=0.] {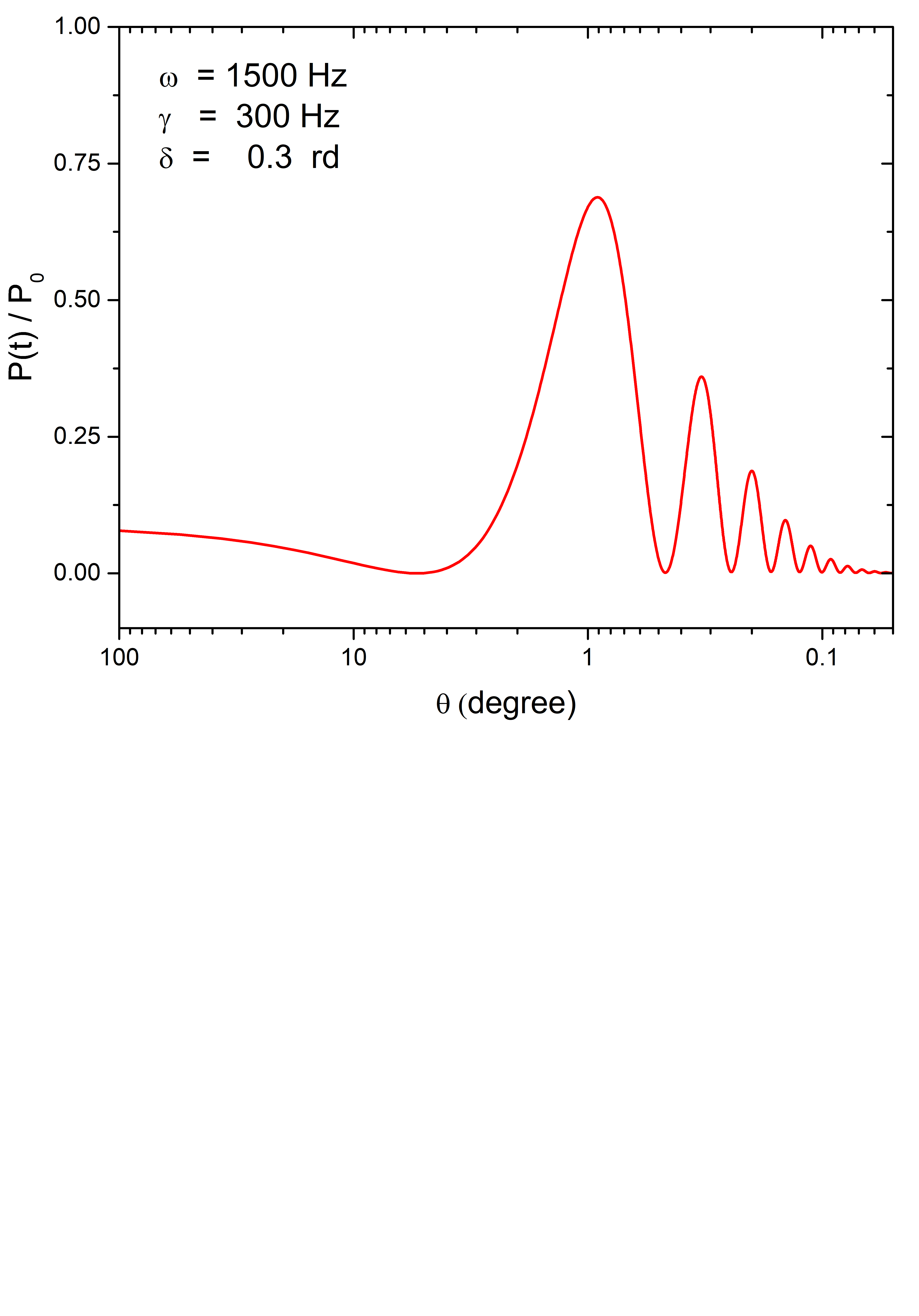}
\vspace*{-4.5cm}
\caption{The red solid line represents the strength of the oscillating microwave radiation density from the superposition of galaxies within the opaque region ($z=0.051$) versus the angular size ($\theta \propto 1/t$). The parameters from the damped oscillation are in the top left corner.
}
\label{damped}
\end{figure} 

Seeing a small angular size in the sky requires a relatively longer time than seeing a larger one, meaning that $\theta \propto t^{-1}$, where $\theta$ is the angular size and 
t is the time.

Fig.~\ref{damped} shows the correlation of the strength of the radiation density  (Eq.~\ref{power3}) and $\theta$, for an oscillation frequency $\omega \cong \omega_0=1.5$ kHz, a damping factor $\gamma = 300$ Hz, a residual $\delta =0.3$, and the $\theta \rightarrow time$ calibration as
\begin{equation}
\left(\frac{\theta}{degree}\right)=0.0011 \left(\frac{t}{s}\right)^{-1},
\end{equation}
An inspection of Fig.~\ref{damped} shows that, with an adjustment of the frequency 
$\omega$ and the damping ratio $\gamma$, the TL model can reproduce up to six oscillations in the angular size range from 1.0 to 0.1 degrees, in agreement with the observed CMB power spectrum. Furthermore, the most prominent and broad peak is the so-called baryonic acoustic oscillation (BAO) in the LCDM model. This ``BAO'' peak is also reproduced in the toy model based on TL.

In addition, Fig.~\ref{damped} also reproduces the shape of the CMB power spectrum for large angular sizes, $\theta \sim 90^{\circ}$, corresponding to the dipole, quadrupole, and higher-order moments.

\section{Sunyaev-Zeldovich effect}
\label{sz_effect}

The spectral distortion of the CMB due to inverse Compton scattering by high-energy electrons in galaxy clusters is known as the Sunyaev–Zeldovich effect (SZ effect) \citep{suny72, suny80, reph95}.

The determination of the Hubble constant using the SZ effect in galaxy clusters, particularly when combined with the gravitational lensing effect of the same clusters, results in a wide distribution of Hubble constant values \citep{cen98}. Similar results were obtained from radio telescope observations of the Comptonization of galaxy clusters, where the Hubble constant ranges from 24 to 76 $ms^{-1}Mpc^{-1}$.
 Comments on these observations can be found in \cite{reph95} and the references therein. One possible explanation for this result is the hypothesis that the CMB photons are not of cosmological origin.

Even if the galaxy cluster redshift is $z \gtrsim 0.051$, the flux of scattered photons from high-energy electrons in the cluster can perturb the opaque region within the optical depth per Hubble radius,  $y_{max}=(nA\; H/c)^{-1}=0.05$. The effect is a decrease in the CMB intensity below 218 GHz and an increase in intensity above 
218 GHz, as observed in the distorted CMB intensity due to the SZ effect in clusters \citep{carl02}.

The differential observational method for searching for the SZ effect in a galaxy cluster, widely used in radio telescope observations, involves choosing an angular gap that includes the galaxy cluster. The intensity measurement outside the angular gap is subtracted from the intensity measurements toward the cluster \citep{reph95}.

To determine the CMB's spectral distortion due to the thermal SZ effect within the TL framework, we use a variation of the above subtraction method. For example, the calculated spectral intensity for the cluster's optical path per Hubble radius $y_{max}>0.05$ is subtracted from the intensity determined for $y_{max}=0.05$ (blackbody).

\begin{figure}[]
\vspace*{-0.0cm}
\hspace*{-0.5cm}
\centering
\includegraphics[clip,width=0.5
\textwidth,height=0.6\textheight,angle=0.] {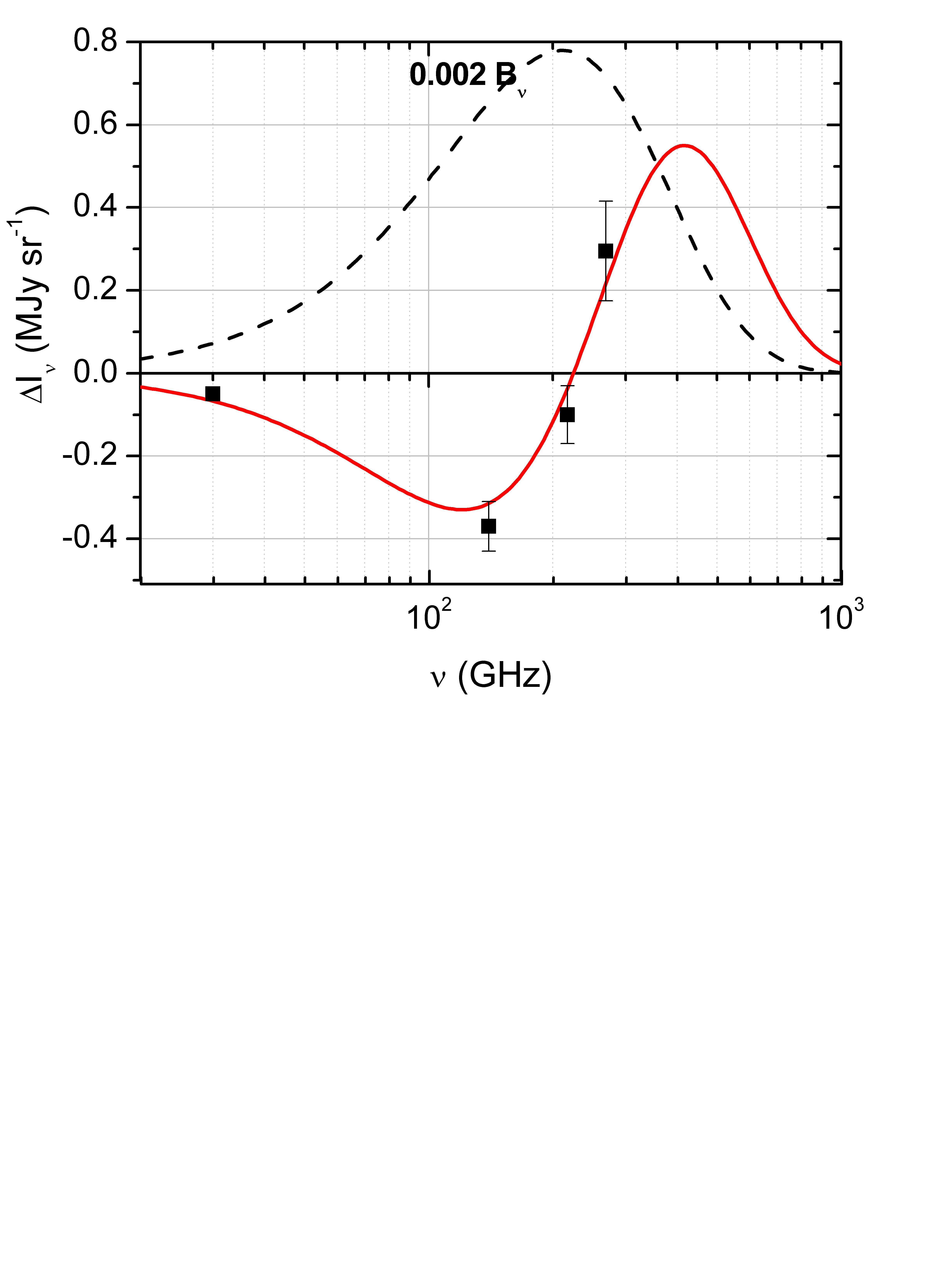}
\vspace*{-7.0cm}
\caption{(Red solid curve) represents distorted CMB spectrum by the 
SZ-effect on the Abell 2163 cluster predicted by the TL model.
(Black dash curve) represents the spectral intensity calculated for an optical depth per Hubble constant of $y_{max}=0.05$ (blackbody), multiplied by 0.002. 
(Data black square points) at 30 GHz from \citep{laro02},  at 140, 218, and 270 GHz from \citep{holz97}.
}
\label{D_sz_cmb}
\end{figure}

The redshift value of the Abell 2163 cluster (z=0.2) suggests an optical path per Hubble radius of $y_{max}=0.182$. With this value, the contribution to the microwave radiation intensity comes from galaxies within a radius of $R_{max}=873.6$ Mpc, and the thermally distorted intensity due to the SZ effect is

\begin{equation}
\Delta I_{\nu}=\eta y \left[(I(ymax=0.182)-I(ymax=0.05)\right].
\end{equation}
Where parameters values to the Abell 2163 cluster are \citep{carl02}: the $y=3.56\times 10^{-4}$ is the Compton parameter, relating the thermal electronic pressure energy along the line of cluster sight \citep{suny72,suny80,zhan04}, $\eta$ is the ratio 
$m_ec^2/kT_e=34.5$, note that the product $\eta y$ is the cluster optical depth to Compton scattering ($\tau_e$)
\begin{equation}
\eta y = \tau_e= \int n_e \sigma_T dl,
\end{equation}
where $n_e$ is the electron number density, $\sigma_T$ is the Thomson cross-section, and the integration
is along the line of sight (size of cluster).

Finally, the $I(ymax=0.182)$ and $I(ymax=0.05)$ are the spectral intensities (see Eq.~\ref{TL_radiance}) calculated for two optical depths per Hubble radius. The last one is like a blackbody intensity. This difference of intensities allows us to have the frequency dependence of the $\Delta I_{\nu}$.

Fig.~\ref{D_sz_cmb} (red solid curve) shows a preliminary prediction of the TL model for the CMB spectrum distortion by the SZ-effect in the Abell 2163 cluster $\Delta I_{\nu}$ versus the observed frequency $\nu_{obs}=\nu e^{\bar{s}}$, where $e^{\bar{s}}=1.2$ is the correction to the frequency (that is shifted) due to collisions. The variable $s=\log(\nu'/\nu)$ is introduced to express the resulting scattering in terms of the logarithmic frequency shift caused by a scattering \citep{suny80,reph95,birk99}.

The black dash curve represents the spectral intensity calculated for an optical depth per Hubble constant of $y_{max}=0.05$ (blackbody), multiplied by 0.002, 
and the data black square points are: at 30 GHz from \citep{laro02},  at 140, 218, and 270 GHz from \citep{holz97}.

In addition to the thermal SZ effect, there is the kinetic SZ effect due to peculiar cluster velocities \citep{suny72}. However, in most cases, the kinetic spectral distortion is about 0.1

On the other hand, the Planck catalog of SZ sources \citep{planck14} comprises 861 identified or confirmed galaxy clusters, most of which were observed by X-ray space observatories and optical ground-based telescopes. Only nine percent of these clusters were detected by the SZ-finder algorithm, and 366 were classified as cluster candidates. A total of 813 Planck clusters have measured redshifts ranging from 
$0.01 \leq z \leq 1.00$, with one-third of the clusters located above z = 0.3.

Additionally, the SZ sources provided by the Planck collaboration with external validation total six (all with known redshifts) \citep{planck14}. The average redshift of these SZ sources is  z=0.177.


\section{Foregrounds}

The CMB is considered the afterglow of the Big Bang, and to detect it as a genuine cosmological signal, the noise in which it is embedded must be removed. In this case, the noise consists of the microwave radiation emitted primarily by the Milky Way.

The term foreground refers to all radiation between the surface of last scattering and the detector. This means that the primary contribution to the foreground is the radiation from the Milky Way. Therefore, an accurate understanding of the foreground emissions is essential for removing them and isolating the cosmological signal.

However, the TL model suggests another scenario for the CMB spectrum. Within the TL model, the CMB results from the superposition of microwave radiation from galaxies ``within an opaque region", with redshifts up to $z=0.051$, traveling distances up to 240 Mpc, and losing up to 5\% of their energy. This new scenario aligns perfectly with the CMB spectrum.

For us, foreground radiation demonstrates that galaxies are sources of microwave radiation. In other words, galaxies emit microwaves that propagate across the universe. A fraction of this radiation reaches Earth.

The local galaxy group (including the Milky Way) is dominated by spiral galaxies, which are rich in gas and dust. This explains why the thermal dust foreground is predominant around and above 100 GHz. Recall 2014, when the BICEP2 collaboration claimed to have observed large-scale B-mode polarization in the CMB \citep{ade14}. However, its origin was later determined to be thermal dust emission from interstellar dust grains.

However, rich galaxy clusters, such as the Virgo cluster, typically have a high concentration of elliptical galaxies near their center. Some of these elliptical galaxies are giant, while spiral galaxies are few and mostly found on the outskirts of the clusters. Ellipticals contain comparatively little gas and dust \citep{trem07} and account for 60\% of observed galaxies. Additionally, there is a fraction (20\%) of irregular galaxies with very little dust.

Furthermore, there are other sources of foreground radiation, most of which predominate in the low-frequency region, with their contribution decreasing as the frequency increases \citep{dick16} for a review. These include galactic free-free radiation emitted by free galactic electrons interacting with ions, synchrotron radiation from relativistic cosmic electrons, and spinning dust radiation from rotating interstellar grains and molecules in the galaxy.

These are the so-called galactic foregrounds (due to the Milky Way). However, there are also foregrounds from extragalactic sources, such as radio galaxies and sub-mm/IR galaxies \citep{toff94,dezo95,gawi97}. In most cases, the discussion focuses on the foreground from nearby galaxies. However, within the TL model, galaxies with a redshift up to 
$z=0.051$ (which includes millions of galaxies) contribute to the CMB.

Moreover, the foreground is complex and difficult to quantify.
There is a large discrepancy in the foreground data from the Planck satellite \citep{adam16} and WMAP \citep{benn13}. For instance, 
between 60–70 GHz, the total foreground at Planck is almost equal to the CMB signal, while at WMAP, the total foreground is about four times lower than the CMB signal.

The galactic foregrounds appear complex, but their total contribution is comparable to the CMB signal at the Planck detector \citep{adam16}. The discrepancy occurs at high frequencies, likely due to detector localization in a region rich in dust. Galactic dust (in addition to emitting microwaves) can absorb and scatter light from distant galaxies. This can lead to incomplete sky maps, such as the Sloan galaxy map (see Fig.~\ref{sloan2}), which shows dark regions due to significant dust.

\section{CMB anomalies}
\label{anomalies}

\subsection{Axis of Evil in the CMB}

At the beginning of the 21st century, U.S. President George Bush referred to several countries, mostly located in the Middle East, as the ``Axis of Evil''. This term was later used by physicists Land and Magueijo \citep{land05} to describe an anomaly observed in data from NASA’s Wilkinson Microwave Anisotropy Probe (WMAP), which shows anisotropic variations in the temperature of the Cosmic Microwave Background (CMB) radiation, aligned with local structures.

According to the cosmological principle, the universe on a large scale is homogeneous and isotropic. In the first approximation, this is observed in the CMB data obtained by WMAP, with temperature variations of approximately $10^{-5}$ degrees Kelvin appearing random. However, an analysis of the temperature variations in higher multipoles reveals that the axes of the quadrupole and octupole moments are parallel and aligned with the ecliptic plane and the direction of the solar system, indicating a preferential direction. This behavior is in tension with the Copernican principle, which is a cornerstone of the standard cosmological model.

The origin of this alignment ranges from a simple statistical fluctuation to more exotic scenarios. However, the ``Axis of Evil'', first observed in the WMAP data, has been confirmed by Planck data.

Considering that there is a tremendous concentration of galaxies near the Milky Way, such as the Virgo Supercluster, one hypothesis is that this peculiar orientation arises from it. About 60\% of the Virgo Supercluster lies within the opaque region, $z \leq 0.051$, where the microwave spectrum follows the blackbody type.

If this is the case, the Axis of Evil could be a signature of the Zwicky drag hypothesis \citep{krag17}, i.e., the gravitational drag of photons by the higher concentration of nearby galaxies, such as those in the Virgo Cluster.

\subsection{North-South asymmetry}

The temperature sky map of the CMB has revealed a statistically significant north-south ecliptic asymmetry \citep{bern08}, with the southern hemisphere slightly hotter than the northern hemisphere. This is in tension with the standard cosmological model, where the fundamental assumption is an isotropic and homogeneous universe.

Some consider the north-south asymmetry to be a statistical fluctuation. However, it was observed in both WMAP and Planck. In other words, the CMB map from ESA's Planck spacecraft improved on WMAP's results but did not resolve the origin of the anomalies. Planck confirmed all of the CMB anomalies first detected by WMAP.

In the TL picture, an excess of galaxies forming clusters in the southern hemisphere and within the opaque region ($z \leq$ 0.051) can explain this asymmetry because galaxy clusters are hot structures that emit X-rays \citep{wilk22}. The existence of a gravitational effect in the CMB, known as the Sachs-Wolfe Effect \citep{sach67}, provides evidence that the distribution of matter (galaxies) in the universe is not uniform.

\subsection{Sachs-Wolfe effect}

The gravitational redshift of CMB photons is known as the Sachs-Wolfe (SW) effect \citep{sach67}. This effect is predominantly responsible for the fluctuations in the CMB on larger angular scales.

We want to remember that Zwick already 1930 postulated the photons gravitational drag hypothesis \citep{krag17}, as a mechanism for photons energy loss hypothesis (see sec.~\ref{drag}).

In addition, the gravitational redshift of photons between the surface of the last scattering and the detector is known as the integrated Sachs-Wolfe (ISW) effect, which depends on the change in the gravitational potential as photons of the CMB pass through a potential well.

The ISW effect is detected in the Planck data at a significance level of only $3\sigma$ \citep{ade16}, which is similar to the detection level achieved from galaxy data. These photons are difficult to reconcile as photons from the CMB and are considered to be of non-cosmological origin. This behavior of the integrated SW effect reinforces the TL model, in which all CMB photons are of non-cosmological origin.

Furthermore, the detection in the Planck data of the integrated SW effect with low confidence (3.2 $\sigma$) \citep{krol22}, in galaxies spanning the redshift range $0 < z < 2$, constrains the so-called ``late integrated Sachs-Wolfe effect." At low redshifts, according to the $\Lambda$CDM model, the universe should be dominated by dark energy, although these potentials do evolve, subtly changing the energy of photons passing through them.

\section{Discussions and Conclusions}
\label{conclusions}

In this paper, we present a toy model based on the TL hypothesis as an alternative explanation for the origin and nature of the CMB.
The optical depth per Hubble radius (for microwaves) predicted by the TL model is large enough to define an opaque region, within this region, the galaxy's microwave radiation superposition behaves like a blackbody, this region
corresponds to a sphere with a radius of 
$R_{max} =240\;Mpc(z=0.051)$, centered on the Milky Way. The microwave radiation emitted by galaxies under these conditions can be identified as the CMB radiation, with an intensity consistent with the COBE FIRAS data. Additionally, the TL model can estimate the size of the CMB temperature fluctuations, which agrees well with the CMB's power spectrum and offers an alternative description of the Sunyaev-Zeldovich effect.

The interpretation of the CMB within the framework of the TL model provides an explanation for the observed anomalies at WMAP and confirmed by Planck, such as the ``axis of evil'' \citep{land05}, an anisotropic variation in the CMB temperature aligned with local structures, and the statistically significant north-south ecliptic asymmetry in the CMB's temperature \citep{bern08}. These anomalies challenge the cosmological principle, which states that the spatial distribution of matter in the universe is uniformly isotropic and homogeneous when viewed on large scales. This principle is a main foundation of the LCDM model.

Even before the JWST dataset, the LCDM model had difficulty correlating the projected angular separations between bright galaxies in a cluster and the cluster's redshift, thus constraining the LCDM model. However, the TL model, under the assumption of a static universe, could fit the data well, without the need for hypothetical entities.

Furthermore, the deep sky images taken by JWST revealed galaxies that were already structured when the universe was only $\sim 0.35$ Gyr old (when the universe was a baby), some of them containing a reasonable amount of carbon \citep{deug24}. These galaxies would therefore be older than the universe itself. These observations constrain Big Bang Nucleosynthesis (BBN), the primordial production of nuclei other than those of the lightest isotopes of hydrogen, helium, and lithium, which are also foundational to the LCDM model.


\section{Acknowledgments}

I thank the Physical Institute of the Universidade Federal Fluminense, Niterói, Rio de Janeiro, Brazil for providing all the conditions necessary to carry out this work,
and the HST, SSDSS, COBE, WMAP and Planck
for their open data policy.

\newpage

\bibliography{tired_light}{}

\begin{thebibliography}{}
\expandafter\ifx\csname natexlab\endcsname\relax\def\natexlab#1{#1}\fi
\providecommand{\url}[1]{\href{#1}{#1}}
\providecommand{\dodoi}[1]{doi:~\href{http://doi.org/#1}{\nolinkurl{#1}}}
\providecommand{\doeprint}[1]{\href{http://ascl.net/#1}{\nolinkurl{http://ascl.net/#1}}}
\providecommand{\doarXiv}[1]{\href{https://arxiv.org/abs/#1}{\nolinkurl{https://arxiv.org/abs/#1}}}

\bibitem[{Adam {et~al.}(2016)Adam, Ade, Aghanim, Alves, Arnaud, Ashdown,
  Aumont, Baccigalupi, Banday, Barreiro, {et~al.}}]{adam16}
Adam, R., Ade, P.~A., Aghanim, N., {et~al.} 2016, Astronomy \& Astrophysics,
  594, A10

\bibitem[{Ade {et~al.}(2014{\natexlab{a}})Ade, Aghanim, Armitage-Caplan,
  Arnaud, Ashdown, Atrio-Barandela, Aumont, Aussel, Baccigalupi, Banday,
  {et~al.}}]{planck14}
Ade, P.~A., Aghanim, N., Armitage-Caplan, C., {et~al.} 2014{\natexlab{a}},
  Astronomy \& Astrophysics, 571, A29

\bibitem[{Ade {et~al.}(2014{\natexlab{b}})Ade, Aikin, Barkats, Benton,
  Bischoff, Bock, Brevik, Buder, Bullock, Dowell, {et~al.}}]{ade14}
Ade, P.~A., Aikin, R.~W., Barkats, D., {et~al.} 2014{\natexlab{b}}, Physical
  Review Letters, 112, 241101

\bibitem[{Ade {et~al.}(2016)Ade, Aghanim, Arnaud, Ashdown, Aumont, Baccigalupi,
  Banday, Barreiro, Bartolo, Basak, {et~al.}}]{ade16}
Ade, P.~A., Aghanim, N., Arnaud, M., {et~al.} 2016, Astronomy \& Astrophysics,
  594, A21

\bibitem[{Alpher {et~al.}(1948)Alpher, Bethe, \& Gamow}]{alph48}
Alpher, R.~A., Bethe, H., \& Gamow, G. 1948, Physical Review, 73, 803

\bibitem[{Alpher \& Herman(1949)}]{alph49}
Alpher, R.~A., \& Herman, R.~C. 1949, Physical Review, 75, 1089

\bibitem[{Assis \& Neves(1995)}]{assi95}
Assis, A. K.~T., \& Neves, M. 1995, Astrophysics and Space Science, 227, 13

\bibitem[{Azzi {et~al.}(2019)Azzi, Farry, Nason, Tricoli, Zeppenfeld, Khalek,
  Alimena, Andari, Bella, Armbruster, {et~al.}}]{azzi19}
Azzi, P., Farry, S., Nason, P., {et~al.} 2019, arXiv preprint arXiv:1902.04070

\bibitem[{Bennett {et~al.}(2013)Bennett, Larson, Weiland, Jarosik, Hinshaw,
  Odegard, Smith, Hill, Gold, Halpern, {et~al.}}]{benn13}
Bennett, C.~L., Larson, D., Weiland, J.~L., {et~al.} 2013, The Astrophysical
  Journal Supplement Series, 208, 20

\bibitem[{Bernui(2008)}]{bern08}
Bernui, A. 2008, Physical Review D—Particles, Fields, Gravitation, and
  Cosmology, 78, 063531

\bibitem[{Birkinshaw(1999)}]{birk99}
Birkinshaw, M. 1999, Physics Reports, 310, 97

\bibitem[{Bondi \& Gold(1948)}]{bond48}
Bondi, H., \& Gold, T. 1948, Monthly Notices of the Royal Astronomical Society,
  108, 252

\bibitem[{Carlstrom {et~al.}(2002)Carlstrom, Holder, \& Reese}]{carl02}
Carlstrom, J.~E., Holder, G.~P., \& Reese, E.~D. 2002, Annual Review of
  Astronomy and Astrophysics, 40, 643

\bibitem[{Cen(1998)}]{cen98}
Cen, R. 1998, The Astrophysical Journal, 498, L99–L101,
  \dodoi{10.1086/311315}

\bibitem[{De~Zotti {et~al.}(1995)De~Zotti, Franceschini, Mazzei, Toffolatti, \&
  Danese}]{dezo95}
De~Zotti, G., Franceschini, A., Mazzei, P., Toffolatti, L., \& Danese, L. 1995,
  Planetary and Space Science, 43, 1439

\bibitem[{Demianski {et~al.}(2012)Demianski, Piedipalumbo, Rubano, \&
  Scudellaro}]{demi12}
Demianski, M., Piedipalumbo, E., Rubano, C., \& Scudellaro, P. 2012, Monthly
  Notices of the Royal Astronomical Society, 426, 1396

\bibitem[{Dicke {et~al.}(1965)Dicke, Peebles, Roll, \& Wilkinson}]{dick65}
Dicke, R.~H., Peebles, P. J.~E., Roll, P.~G., \& Wilkinson, D.~T. 1965,
  Astrophysical Journal, vol. 142, p. 414-419, 142, 414

\bibitem[{Dickinson(2016)}]{dick16}
Dickinson, C. 2016, arXiv preprint arXiv:1606.03606

\bibitem[{Dirac(1938)}]{dira38}
Dirac, P. A.~M. 1938, Proceedings of the Royal Society of London. Series A.
  Mathematical and Physical Sciences, 165, 199

\bibitem[{D’Eugenio {et~al.}(2024)D’Eugenio, Maiolino, Carniani,
  Chevallard, Curtis-Lake, Witstok, Charlot, Baker, Arribas, Boyett,
  {et~al.}}]{deug24}
D’Eugenio, F., Maiolino, R., Carniani, S., {et~al.} 2024, Astronomy \&
  Astrophysics, 689, A152

\bibitem[{Finlay-Freundlich(1954)}]{finl54}
Finlay-Freundlich, E. 1954, Proceedings of the Physical Society. Section A, 67,
  192

\bibitem[{Gabrielli {et~al.}(2004)Gabrielli, Labini, Joyce, \&
  Pietronero}]{gabr04}
Gabrielli, A., Labini, F., Joyce, M., \& Pietronero, L. 2004, Física
  Estatistica para Estruturas Cosmicas (Springer. pág. 377. ISBN
  978-3-540-40745-4)

\bibitem[{Gamow(1948)}]{gamo48}
Gamow, G. 1948, Nature, 162, 680

\bibitem[{Gawiser \& Smoot(1997)}]{gawi97}
Gawiser, E., \& Smoot, G.~F. 1997, The Astrophysical Journal, 480, L1

\bibitem[{Gupta(2023)}]{gupt23}
Gupta, R.~P. 2023, Monthly Notices of the Royal Astronomical Society, 524, 3385

\bibitem[{Hickson \& Adams(1979)}]{hick79}
Hickson, P., \& Adams, P.~J. 1979, Astrophysical Journal, Part 2-Letters to the
  Editor, vol. 234, Dec. 1, 1979, p. L91-L95., 234, L91

\bibitem[{Holzapfel {et~al.}(1997)Holzapfel, Ade, Church, Mauskopf, Rephaeli,
  Wilbanks, \& Lange}]{holz97}
Holzapfel, W., Ade, P.~A., Church, S., {et~al.} 1997, The Astrophysical
  Journal, 481, 35

\bibitem[{Kragh(2017)}]{krag17}
Kragh, H. 2017, Journal of Astronomical History and Heritage, 20, 2

\bibitem[{Krolewski \& Ferraro(2022)}]{krol22}
Krolewski, A., \& Ferraro, S. 2022, Journal of Cosmology and Astroparticle
  Physics, 2022, 033

\bibitem[{Land \& Magueijo(2005)}]{land05}
Land, K., \& Magueijo, J. 2005, Physical Review Letters, 95, 071301

\bibitem[{LaRoque {et~al.}(2002)LaRoque, Carlstrom, Reese, Holder, Holzapfel,
  Joy, \& Grego}]{laro02}
LaRoque, S., Carlstrom, J., Reese, E., {et~al.} 2002, Arxiv preprint
  astro-ph/0204134

\bibitem[{LaViolette(1986)}]{lavi86}
LaViolette, P.~A. 1986, Astrophysical Journal, Part 1 (ISSN 0004-637X), vol.
  301, Feb. 15, 1986, p. 544-553., 301, 544

\bibitem[{L{\'o}pez-Corredoira \& Marmet(2022)}]{lope22}
L{\'o}pez-Corredoira, M., \& Marmet, L. 2022, International Journal of Modern
  Physics D, 31, 2230014

\bibitem[{Lovyagin {et~al.}(2022)Lovyagin, Raikov, Yershov, \&
  Lovyagin}]{lovy22}
Lovyagin, N., Raikov, A., Yershov, V., \& Lovyagin, Y. 2022, Galaxies, 10, 108

\bibitem[{Pecker(1997)}]{peck97}
Pecker, J.-C. 1997, Journal of Astrophysics and Astronomy, 18, 323

\bibitem[{Penzias \& Wilson(1979)}]{penz79}
Penzias, A.~A., \& Wilson, R.~W. 1979, in A Source Book in Astronomy and
  Astrophysics, 1900--1975 (Harvard University Press), 873--876

\bibitem[{Rephaeli(1995)}]{reph95}
Rephaeli, Y. 1995, Annual Review of Astronomy and Astrophysics, Volume 33,
  1995, pp. 541-580., 33, 541

\bibitem[{Riess {et~al.}(2007)Riess, Strolger, Casertano, Ferguson, Mobasher,
  Gold, Challis, Filippenko, Jha, Li, Tonry, Foley, Kirshner, Dickinson,
  MacDonald, Eisenstein, Livio, Younger, Xu, Dahlen, \& Stern}]{ries07}
Riess, A.~G., Strolger, L., Casertano, S., {et~al.} 2007, The Astrophysical
  Journal, 659, 98–121, \dodoi{10.1086/510378}

\bibitem[{Riess {et~al.}(2024)Riess, Scolnic, Anand, Breuval, Casertano, Macri,
  Li, Yuan, Huang, Jha, {et~al.}}]{ries24}
Riess, A.~G., Scolnic, D., Anand, G.~S., {et~al.} 2024, arXiv preprint
  arXiv:2408.11770

\bibitem[{Sachs \& Wolfe(1967)}]{sach67}
Sachs, R., \& Wolfe, A. 1967, The Astrophysical Journal, 147, 73

\bibitem[{Shamir(2024)}]{sham24}
Shamir, L. 2024, Particles, 7, 703

\bibitem[{Skordis \& Z{\l}o{\'s}nik(2021)}]{skor21}
Skordis, C., \& Z{\l}o{\'s}nik, T. 2021, Physical Review Letters, 127, 161302

\bibitem[{SkyServer(2008)}]{sloan4}
SkyServer, S. 2008, {Spectra of stars},
  \url{https://skyserver.sdss.org/dr7/en/proj/basic/spectraltypes/stellarspectra.asp}

\bibitem[{Sunyaev \& Zeldovich(1980)}]{suny80}
Sunyaev, R., \& Zeldovich, I.~B. 1980, In: Annual review of astronomy and
  astrophysics. Volume 18.(A81-20334 07-90) Palo Alto, Calif., Annual Reviews,
  Inc., 1980, p. 537-560., 18, 537

\bibitem[{Sunyaev \& Zeldovich(1972)}]{suny72}
Sunyaev, R.~A., \& Zeldovich, Y.~B. 1972, Comments Astrophys. Space Phys., 4,
  173

\bibitem[{Suzuki \& Fukugita(2018)}]{suzu18}
Suzuki, N., \& Fukugita, M. 2018, The Astronomical Journal, 156, 219

\bibitem[{Toffolatti(1994)}]{toff94}
Toffolatti, L. 1994, arXiv preprint astro-ph/9410037

\bibitem[{Tremblay {et~al.}(2007)Tremblay, Chiaberge, Donzelli, Quillen,
  Capetti, Sparks, \& Macchetto}]{trem07}
Tremblay, G.~R., Chiaberge, M., Donzelli, C.~J., {et~al.} 2007, The
  Astrophysical Journal, 666, 109

\bibitem[{Wilkes {et~al.}(2022)Wilkes, Tucker, Schartel, \&
  Santos-Lleo}]{wilk22}
Wilkes, B.~J., Tucker, W., Schartel, N., \& Santos-Lleo, M. 2022, Nature, 606,
  261

\bibitem[{Zhang {et~al.}(2004)Zhang, Pen, \& Trac}]{zhan04}
Zhang, P., Pen, U.-L., \& Trac, H. 2004, arXiv preprint astro-ph/0402115

\bibitem[{Zwicky(1929)}]{zwic29}
Zwicky, F. 1929, Proceedings of the National Academy of Sciences, 15, 773

\end{thebibliography}
\bibliographystyle{aasjournal}

\end{document}